\documentclass[prl,reprint,aps,epsf,showpacs,superscriptaddress,longbibliography]{revtex4-2}

\usepackage{amsmath}
\usepackage{amssymb}
\usepackage{dsfont}
\usepackage{graphicx}
\usepackage{hyperref}
\usepackage{float}
\usepackage{soul}
\usepackage{wasysym}
\usepackage{bbm}

\usepackage{pgffor}

\setcitestyle{super}

\newcommand{\ket}[1]{\ensuremath{|#1 \rangle}}

\usepackage[dvipsnames]{xcolor}
\usepackage[normalem]{ulem}
\usepackage{comment}

\usepackage{color}
\usepackage{hyperref}

\hypersetup{colorlinks,citecolor=blue,linkcolor=blue,urlcolor=blue}

\usepackage{quantikz}

\begin{document}

\title{Error detection without post-selection in adaptive quantum circuits}

\author{Eli Chertkov}
\email{eli.chertkov@quantinuum.com}
\author{Andrew C. Potter}
\author{David Hayes}
\author{Michael Foss-Feig}
\affiliation{Quantinuum, 303 South Technology Court, Broomfield, Colorado 80021, USA}

\begin{abstract}
Current quantum computers are limited by errors, but have not yet achieved the scale required to benefit from active error correction in large computations.
We show how simulations of open quantum systems can benefit from error \emph{detection}. In particular, we use Quantinuum’s H2 quantum computer to perform logical simulations of a non-equilibrium phase transition using the [[4,2,2]] code. Importantly, by converting detected errors into random resets, which are an intended part of the dissipative quantum dynamics being studied, we avoid any post-selection in our simulations, thereby eliminating the exponential cost typically associated with error detection. The encoded simulations perform near break-even with unencoded simulations at short times.
\end{abstract}

\maketitle

\begin{figure}
    \centering
    \includegraphics[width=0.5\textwidth]{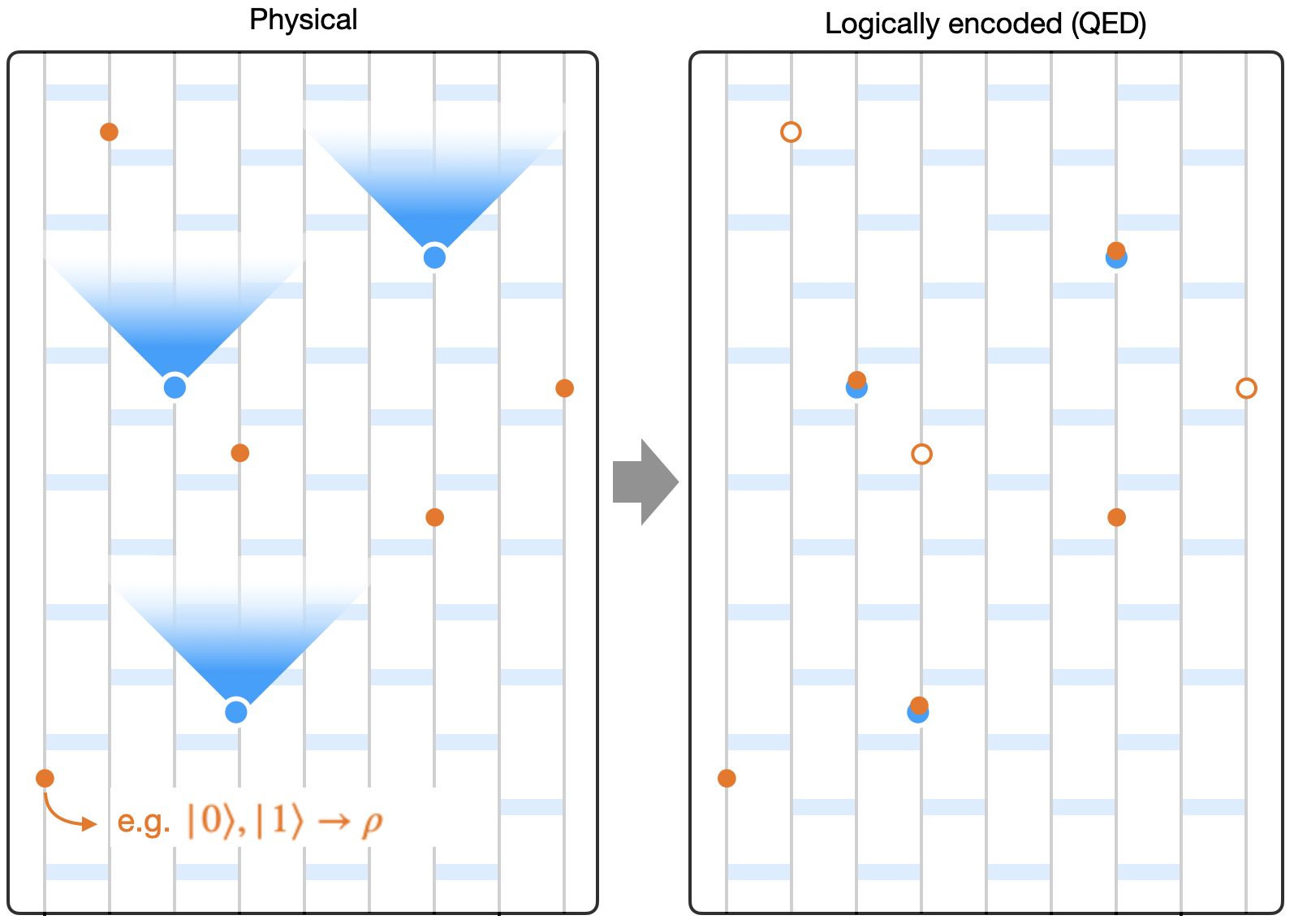}
    \caption{How error detection can be used to remove errors without post-selection in a quantum circuit with random resets (e.g., to a mixed state $\rho$). (\textbf{Left}) In the physical circuit, resets (orange) are randomly inserted and hardware errors (blue) occur randomly, spreading and corrupting the circuit's output. (\textbf{Right}) In the logical circuit encoded into an error detection code, hardware errors can be detected and converted into resets, preventing the spread of errors without needing to discard the circuit realization. Note that the probability of a reset is now increased by the detection events, so the injected resets rate needs to be decreased to compensate.}
    \label{fig:fig1}
\end{figure}

\emph{Introduction---} Large-scale implementations of quantum error correction (QEC) in digital quantum computers are essential for realizing scalable quantum computation. However, QEC requires substantial overheads in quantum resources\cite{Bravyi2005}, creating a large barrier to its application in near-term resource-limited devices. Quantum error detection (QED), which involves using a logical encoding to only detect but not correct errors, has a substantially smaller overhead than QEC and provides some of the same benefits, though is generally not scalable. The standard way of using QED -- post-selecting on circuit realizations with no detected errors -- incurs an exponential run-time overhead \cite{Corcoles2015,Gottesman2016,Vuillot2017,Linke2017,Harper2019,Wang2024,Bluvstein2024,Self2024,Yamamoto2024,He2025,Jin2025,Dasu2025}. In this work, we present and experimentally demonstrate a scalable error detection protocol that has no post-selection overhead when applied to the quantum simulation of certain dissipative quantum circuits (see Fig.~\ref{fig:fig1}).

The dissipative quantum circuit we focus on is a quantum variant of a classical model known as the contact process \cite{Harris1974}. The classical contact process is a stochastic process, similar to a cellular automaton, that describes how a disease spreads in a population. The model has been found to exhibit a non-equilibrium phase transition, a type of phase transition that occurs far from thermal equilibrium \cite{Marro1999,Hinrichsen2000,Odor2004,Hinrichsen2006}, in the directed percolation (DP) universality class \cite{Jensen1999,Hinrichsen2000}. 
Recent studies have quantized the contact process to understand whether quantum fluctuations affect the critical physics, arguing that the quantum models do not follow DP universality \cite{Marcuzzi2016,Carollo2019,Gillman2019,Minjae2021}.
Ref.~\onlinecite{Chertkov2023} studied a dissipative quantum circuit version of the contact process, called the Floquet quantum contact process (FQCP), that contains periodic unitary evolution followed by dissipative mid-circuit random resets. Through a numerical study and an experimental demonstration on Quantinuum's H1 quantum computer, they found that the FQCP appeared to be in the DP universality class \cite{Chertkov2023}. In this work, we study a slightly modified variant of the FQCP model near its critical point.

The phase transition that occurs in the dissipative circuit, also sometimes referred to as an absorbing state transition, also has connections with measurement-induced phase transitions (MIPT) \cite{Skinner2019,Potter2022}. The MIPT is a phase transition that can be detected through the entanglement properties of quantum trajectories in circuits with mid-circuit measurements. Since the MIPT properties are trajectory dependent, studying it on a quantum computer encounters an exponential post-selection overhead. Recently, numerous works have explored absorbing state transitions in quantum circuits, which have random mid-circuit resets and no post-selection overhead, to understand their connections to the MIPT transition \cite{Buchhold2022,Ravindranath2023,ODea2024,Sierant2023,Sierant2023b,Qian2024,LeMaire2023}. Generally, it has been found that quantum circuits exhibiting absorbing state transitions also contain MIPT transitions in their ``active'' phase \cite{ODea2024,Sierant2023,Sierant2023b}. We expect the same to be true in our model.

In this work, we implement the contact process dissipative circuit using the [[4,2,2]] code \cite{Vaidman1996,Grassl1997}, the smallest example of an error detecting code, on Quantinuum's H2 trapped-ion quantum computer \cite{DeCross2024,Moses2023}. Our circuit is adaptive, with hardware errors detected by the [[4,2,2]] code used to trigger the random resets in the circuit. We show that the protocol is scalable and use the adaptive logical circuits to measure quantitatively accurate observables near the phase transition that appear to break-even with the results from physical circuits. 

\begin{figure}
    \centering
    \includegraphics[width=0.5\textwidth]{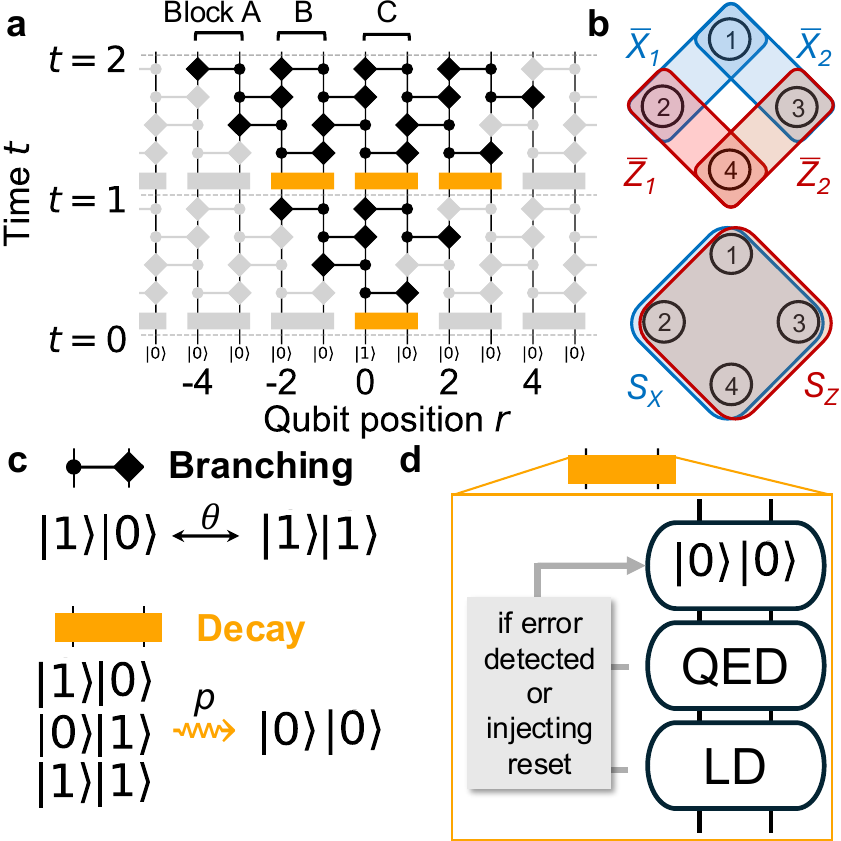}
    \caption{\textbf{a} The dissipative quantum circuit studied in this work, encoded into multiple [[4,2,2]] quantum error detection code blocks. Each qubit shown is a logical qubit and gates are logical gates (the encoding using physical qubits is not shown here). The two-qubit gates are controlled-$R_x(\theta)$ rotation gates. The orange boxes are random two-qubit resets that occur with probability $p$. \textbf{b} The [[4,2,2]] code's logical Pauli operators $\bar{Z}_{1/2},\bar{X}_{1/2}$ and stabilizers $S_{Z/X}$. \textbf{c} The two competing processes in the circuit acting on computational basis states, a branching process that spreads $|1\rangle$ states to other qubits (controlled by $\theta$) and a decay process that causes $|1\rangle$ states to decay to $|0\rangle$ states (at a rate controlled by $p$). \textbf{d} Using an error detection code, the circuit can be made adaptive so that random resets are intentionally injected at random or are triggered by a leakage detection (LD) or error detection (QED) event generated by quantum hardware noise.}
    \label{fig:fig2}
\end{figure}

\emph{Quantum error detection code---} In this work, we use an $[[n,k,d]]=[[4,2,2]]$ code -- a quantum error detection code that encodes $k=2$ logical qubits into $n=4$ physical qubits. Since it is a distance $d=2$ code, it can detect (but not correct) a single error. It is the smallest code capable of protecting quantum information and is the smallest code in the family of $[[n,n-2,2]]$ ``Iceberg'' codes \cite{Chao2018,Self2024}. Each [[4,2,2]] code block has two weight-4 stabilizers $S_Z=Z_1 Z_2 Z_3 Z_4,S_X=X_1 X_2 X_3 X_4$ and four logical operators $\bar{X}_{1,2},\bar{Z}_{1,2}$ implemented by physical weight-2 operators, as shown in Fig.~\ref{fig:fig2}\textbf{b}. Because logical operators are weight-2 (since $d=2$), logical rotations can be implemented (non-fault-tolerantly) using a single $U_{ZZ}(\theta)=e^{-i\theta Z \otimes Z/2}$ gate, which can be implemented natively on Quantinuum hardware \cite{Chertkov2023,Moses2023,DeCross2024}.

In the context of quantum error detection, stabilizers are typically measured one or more times during each shot and measurement results from shots with detected errors are discarded. However, in this work, we do not discard shots and therefore need to consider the logical effects of individual errors. We refer to a Pauli $P$ that triggers (does not trigger) a stabilizer violation as a detectable (undetectable) error; similarly, a Pauli that would flip a logical measurement result is called a logical error (e.g., $X_1 X_2$ is an undetectable logical error and $X_1$ is a detectable logical error if measuring $\bar{Z}_1=Z_1Z_3$).

Operations in quantum error correction and error detection codes can be implemented in many ways, some fault-tolerant (FT) and others not \cite{Gottesman2022}. In the context of the $[[4,2,2]]$ code, a FT operation is one in which any single circuit fault causes a detectable error, i.e., which has a logical fidelity \emph{after post-selection} that scales as $\sim Bp^2$ when physical errors happen with probability $\sim A p$ for small $p$. For a FT operation, there is a pseudo-threshold $p_c = A/B$, which indicates the value of $p$ below which the logical operation outperforms the physical one. In the [[4,2,2]] code, intrablock $\overline{CNOT},\bar{H}\otimes\bar{H}$ gates as well as interblock pairs of transversal CNOT gates are FT \cite{Gottesman2016,Harper2019}.

Unfortunately, implementing a universal FT gate set necessarily requires significant overhead, particularly for performing small-angle rotation gates required for quantum simulation~\cite{Eastin2009}. However, QEC codes can still suppress errors for operations that are non-FT.
If a logical operation is not FT and fails with probability $\sim Cp$, that operation can in principle still outperform the physical one if $C < A$. 
This inequality tends to be reversed for high-distance QEC codes whose logical operations have many more gates than their unencoded physical versions. By contrast, in the [[4,2,2]] code, the logical two-qubit rotation $e^{-i\theta \bar{Z}_1  \bar{Z}_2 /2}=e^{-i\theta Z_2 Z_3/2}$ can be realized as a single native $U_{ZZ}(\theta)$ whose error model has been shown \cite{Haghshenas2025} to contain very low probability of generating undetectable logical errors ($X\otimes X,Y \otimes Y, Z \otimes Z$ errors). This means that this gate, even though it is non-FT, effectively has $C \ll A$ making the non-FT gate (potentially) better than the physical gate. However, non-FT gates can spread errors pathologically, potentially converting detectable logical errors on one code block into undetectable logical errors on others (see Supplement~I.C for an example).

For the remainder of the paper, we drop the overline notation for logical operators and states. Operators and states are at the logical level unless stated otherwise.

\emph{Quantum contact process model---} 
Using the [[4,2,2]] code, we encode the dynamics of a one-dimensional dissipative quantum circuit analogous to the classical contact process \cite{Harris1974} that undergoes an absorbing state phase transition \cite{Hinrichsen2000,Hinrichsen2006}. The contact process circuit (see Fig.~\ref{fig:fig2}\textbf{a}) is a time-periodic quantum channel $\mathcal{E}(\rho) = \left(\mathcal{U}_\theta \mathcal{R}_p \right)^t$ made of unitaries $\mathcal{U}_\theta(\rho)=U_\theta\rho U_\theta^\dagger$ defined by an angle $\theta$ and random reset quantum channels $\mathcal{R}_p$ defined by the probability $p$, similar to the model introduced in Ref.~\onlinecite{Chertkov2023}. The unitary operator $U_\theta$ contains four alternating layers of controlled-rotation gates $CR_x(\theta)=e^{-\frac{i}{4}\theta(1+Z)\otimes X}$, and the random reset channel $\mathcal{R}_p=\prod_{k=1}^{L/2}\mathcal{R}_{2k,2k+1}$ consists of a single layer of \emph{two-qubit} random reset channels $\mathcal{R}_{j,j+1}(\rho^{(2)})=(1-p)\rho^{(2)}_{j,j+1} + p|00\rangle\langle 00|_{j,j+1}$. Importantly, both $\mathcal{U}_\theta$ and $\mathcal{R}_p$ preserve the $\ket{0\cdots 0}$ product state, which is called the absorbing state.

The competition between the unitary evolution, which causes spreading of $\ket{1}$ (active) states among qubits, and the dissipation, which causes decay to the $\ket{0}$ (inactive) state, (see Fig.~\ref{fig:fig2}\textbf{c}) results in an absorbing state transition at a critical $p_c$ in the directed percolation universality class. In our experiments, we examine the propagation of a single active state in the center of the chain (see Fig.~\ref{fig:fig2}\textbf{a}). For $p<p_c$, the model is in the ``active'' phase  and active sites proliferate over time. For $p > p_c$, the model is in the ``absorbing'' phase and active sites decay away so that the system falls into the absorbing state, which it cannot escape from. At the $p=p_c$ critical point, active sites still proliferate, though in a self-similar fractal way with power-law scaling governed by DP critical exponents \cite{Hinrichsen2000,Hinrichsen2006,Jensen1999}. In studies of similar continuous-time models, it was argued \cite{Marcuzzi2016,Carollo2019,Gillman2019} that quantum effects modified the critical exponents from the DP values. In classical numerical simulations for this model (see Supplement~II) and its previous version \cite{Chertkov2023}, we find the exponents are consistent with DP.

In this work, we use a contact process model with two-qubit resets in order to match the two-qubit code block of the [[4,2,2]] code. Since the number of qubits involved in the reset does not alter any of the universal features of the model such as spatial locality, dimensionality, or structure of the absorbing state, we expect the phase transition in this model to fall into the same universality class as that of its single-qubit-reset version. As described in Supplement~I.B, logical resets, error detection, and measurements are implemented fault-tolerantly. However, to avoid the prohibitive resource overhead of FT magic state injection \cite{Bravyi2005}, we implement the non-Clifford two-qubit gates (both within a code block and between code blocks) in a non-FT way, which ultimately limits the accuracy of our final results. The state preparation, error detection, and measurement require ancilla qubits.
Because H2 implements gates in batches of four qubits, we include only four additional ancilla qubits, which we reset and reuse for each of the relevant logical operations.

\begin{figure*}
    \centering
    \includegraphics[width=\textwidth]{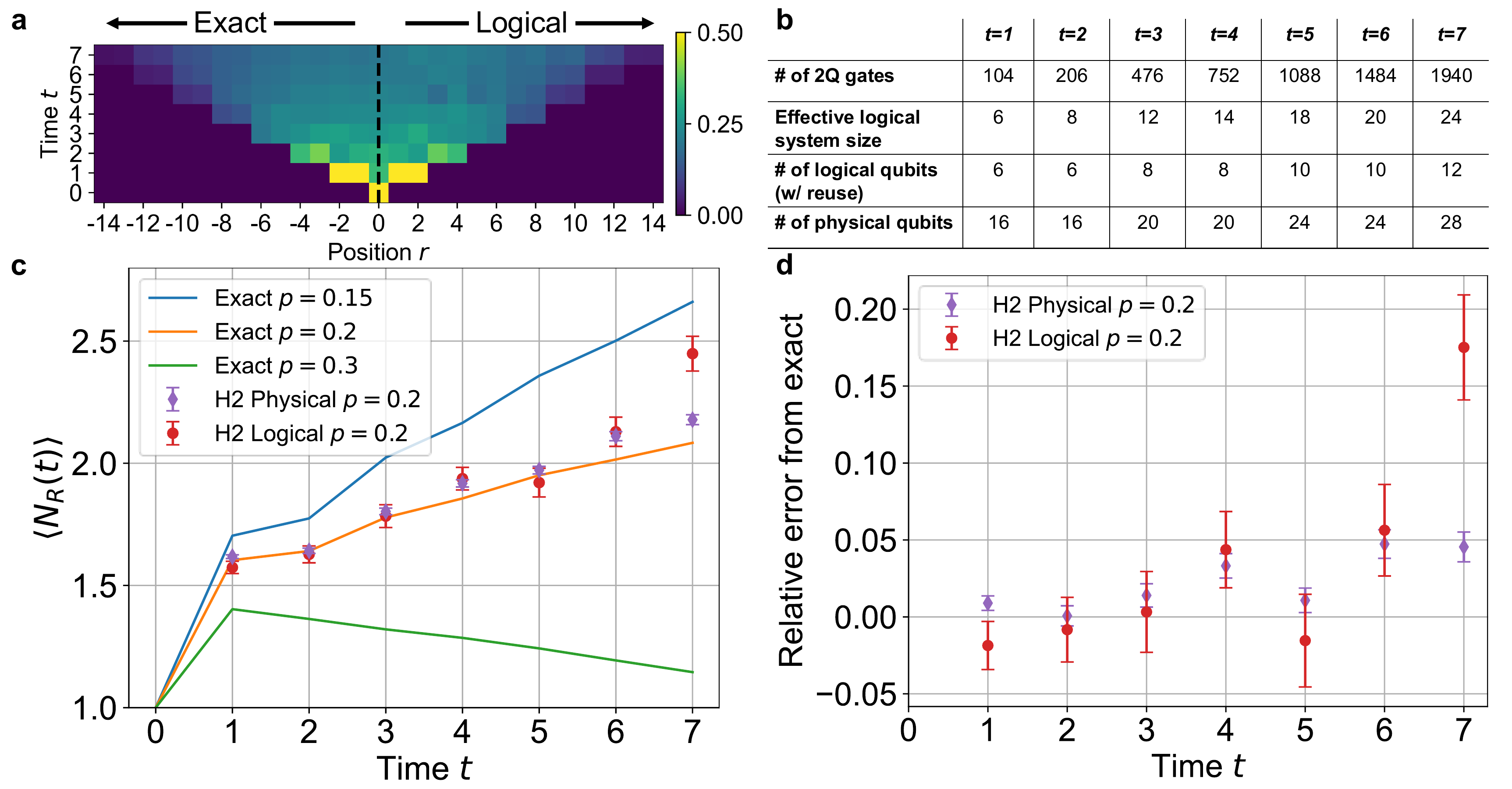}
    \caption{\textbf{a} The active site density profile $\langle n(r,t)\rangle$ obtained from the logical circuit run on H2 compared with the exact profile computed from classical numerics (reflected about $r=0$). \textbf{b} The two-qubit (2Q) gate and qubit overheads of the logical circuits with and without qubit-reuse. \textbf{c} The experimental results obtained on H2. The total number of active sites on the right half of the system versus time, comparing physical (purple diamonds) and logical results (red circles) against classical exact numerics (solid lines). \textbf{d} The relative error of the observable in \textbf{c} from the exact result for the logical and physical circuits.}
    \label{fig:fig3}
\end{figure*}

We use qubit-reuse compilation \cite{DeCross2022,kim2017,fossfeig2020,Barratt2021,Chertkov2022,Niu2021,Zhang2022} at the logical level, which allows us to reduce the qubit resources required to execute the circuit by opportunistically resetting and reusing logical code blocks. Without qubit-reuse the largest logical circuits that we run would require $52$ physical qubits (including four ancilla). Instead, with qubit reuse, we ran them on the H2 quantum computer using only $28$ physical qubits (see Fig.~\ref{fig:fig3}\textbf{b}).

\emph{Quantum circuits that adapt to noise---} We avoid the exponential post-selection penalty of quantum error detection by using an adaptive quantum circuit whose dynamics are controlled by the mid-circuit error detection measurements. As shown in Fig.~\ref{fig:fig2}\textbf{a}, in each time step of the contact process circuit (every four layers of gates) a layer of random two-qubit resets occurs that resets each code block with probability $p$. In our logical circuit, we perform stabilizer measurements and leakage detection measurements \cite{Stricker2020,Moses2023,Haghshenas2025} of each code block right before the random resets are applied. As shown in Fig.~\ref{fig:fig2}\textbf{d}, if a stabilizer violation or leakage error (or detection gadget error) is detected then that code block is reset to $\ket{00}$. The probability of an error being detected and converted to a reset $p_{\textrm{detect}}(r,t)$ is a function of the gate error model, circuit structure, and memory error induced by ion transport. In addition to the ``detected resets'' directly correlated to the error detection events, we also inject resets randomly with probability $p_{\textrm{inject}}$ by generating random numbers and conditioning the random resets based on their values. We generate the random numbers using the quantum computer itself before the start of the calculation, as done in Ref.~\onlinecite{Chertkov2023}.

Given a target reset rate $p$ for our quantum circuit, with this protocol $p_{\textrm{detect}}(r,t) \leq p$ must hold for all space-time points. One can in principle use this protocol when $p_{\textrm{detect}} > p$ by discarding enough shots to lower the detection probability, but this creates an exponential post-selection overhead that the protocol was designed to avoid so we do not consider this here. We run this adaptive protocol in two steps: a calibration run and a main run. In the calibration run, we do not inject any resets but only allow for resets generated by detection events and thereby obtain a measure for $p_{\textrm{detect}}(r,t)$. In the main run, we inject resets at the rate
\begin{align}
    p_{\textrm{inject}}(r,t) = \frac{p-p_{\textrm{detect}}(r,t)}{1-p_{\textrm{detect}}(r,t)} \label{eq:pinject}
\end{align}
chosen so that probability of a reset occurring from either a detection or injection event is $p$. In this protocol, we are assuming that detected resets are uncorrelated with one another and that injected resets do not appreciably affect the error detection rate, which empirically appears to hold in our results.

When a detected reset occurs in a code block, the corrupted logical state is replaced by an uncorrupted $\ket{00}$ state, thereby preventing the further impact of that error in the code block without any post-selection. However, due to the non-FT interblock two-qubit gates in the model, the detectable error that caused the reset can propagate into \emph{undetectable} logical errors on other code blocks that are left unmitigated.

\emph{Experimental results ---} We implement the above-described contact process circuits on Quantinuum's H2-2 trapped-ion quantum computer, which has high-fidelity mid-circuit measurements, resets, and classical feed-forward capabilities -- crucial features for executing these adaptive circuits. 

We time-evolve an initial product state $\ket{\psi(0)} = \ket{0\cdots010\cdots 0}$, with a single active site at position $r=0$ for times $t=1,2,\ldots,7$ under the contact process dynamics with parameters $\theta=3\pi/4$ (same as Ref.~\onlinecite{Chertkov2023}, far from the Clifford point at $\theta=\pi/2$) and $p=0.2$. This parameter is near the critical point of the model determined from classical numerics (see Supplement~II.B). We perform two sets of experiments, one with the adaptive logical circuit encoded in the [[4,2,2]] code and one with the physical unencoded circuit. The physical circuit only has injected resets and also uses qubit-reuse. 
For each time $t$, we measure the spatial profile of the active-site density $n(r)=|1\rangle\langle1|_r=(I+Z_r)/2$ and the total-number of active sites on the right-half of the system $N_R=\sum_{r\geq 0}n(r)$, which contain information about the critical spreading of active sites. For the logical circuits we run 1,000 shots at each time step for both the calibration run and main runs; for the physical circuits we run 10,000 shots at each time step. 
The qubit number and two-qubit gate overheads of the logical circuit are listed in Fig.~\ref{fig:fig3}\textbf{b}.

Fig.~\ref{fig:fig3}\textbf{a} shows a heatmap of the active-site density profile obtained from the logical circuits (right) compared with exact classical density matrix simulations (left). 
Visually, we see that the agreement is quite good with the logical circuit clearly displaying quantitatively accurate sub-ballistic spreading of active sites. Fig.~\ref{fig:fig3}\textbf{c} shows the number of active sites versus time measured for the logical and physical circuits run on H2 compared with exact numerics. All error bars reported are standard errors obtained from bootstrap resampling using 200 resamples. Below the transition (blue) $\langle N_R(t)\rangle$ should grow linearly, above the transition (green) $\langle N_R(t)\rangle$ should decay to zero, and near the transition (orange) $\langle N_R(t)\rangle \sim t^\Theta$ with DP exponent\cite{Jensen1999}  $\Theta=0.313686(8)$. Fig.~\ref{fig:fig3}\textbf{d} shows the relative error in this observable between logical and physical from exact. We see that for times $t=1,2,3,5$ the logical and physical circuits agree within error bars with each other and with the numerically-exact curve. For $t=4,6$, both physical and logical results agree within error bars with each other, but differ by $\gtrsim 2$ standard-errors from the exact result. Finally, at $t=7$ the logical result differs significantly more from exact than the physical result.
We attribute this strong deviation at $t=7$ to accumulation of memory/idling errors: Since H2-2 can execute two-qubit gates in batches of at most four, and the number of gates per circuit layer grows linearly with simulation time $t$ (see Fig.~\ref{fig:fig2}\textbf{a}), the delay time and memory error per circuit layer correspondingly grows as $\sim t$. 
This ultimately sets a practical limit on the maximal achievable simulation time depth $t$.
Note that while the physical circuits also have memory errors that scale as $\sim t$, the prefactor is much larger in the logical circuit because logical operations require many more physical gates and thereby many more ion transport operations. Ultimately, we see that our results are consistent with the logical circuit breaking even with the physical circuit for $t\leq 6$.

\begin{figure}
    \centering
    \includegraphics[width=0.5\textwidth]{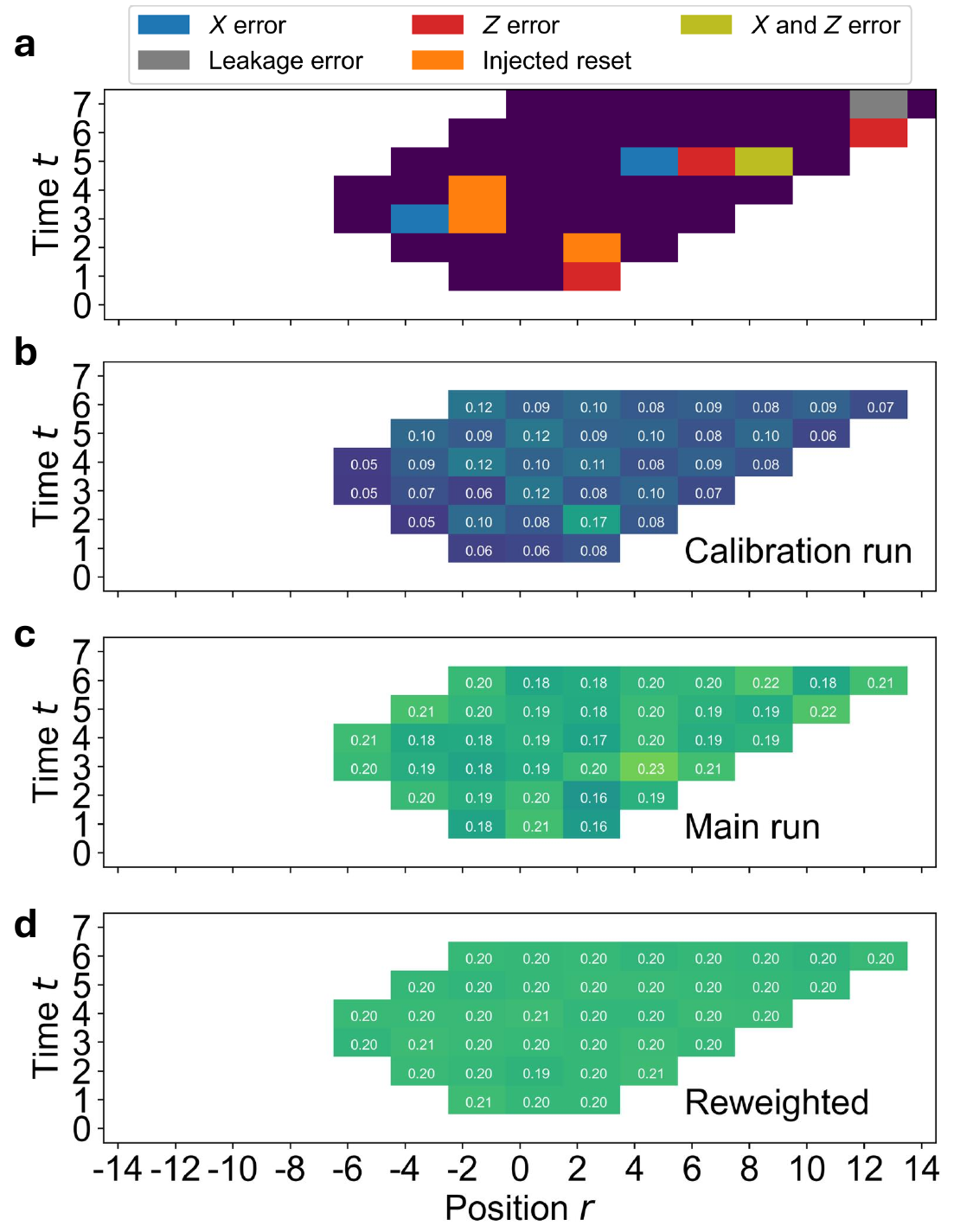}
    \caption{\textbf{a} The space-time record of error detection events and injected resets for one shot of a $t=7, p=0.2$ circuit run on H2. \textbf{b} The average space-time probabilities of detecting an error in the calibration run. \textbf{c} The average space-time probabilities of resetting a code block in the main run, which includes injected resets as well as detected resets. \textbf{d} The average space-time probabilities of resetting a code block after using the reweighting scheme.}
    \label{fig:fig4}
\end{figure}

In addition to the observables, we also inspect the reset rates obtained in the adaptive logical circuits to verify that resets are properly sampled. Fig.~\ref{fig:fig4}\textbf{a} shows an example record of error detection outcomes obtained in one shot on the H2 quantum computer. In this record, we can see the space-time locations of both detected and injected resets.
For the calibration run at time $t=7$, if we average over all shots we obtain the results in Fig.~\ref{fig:fig4}\textbf{b}, which shows the average error detection rate (for all types of errors) at all space-time points. The resulting reset rates are visibly inhomogeneous in space and time, predominately due to memory error accumulated during ion transport in the qubit-reuse compilation. 
In the supplement, we discuss the various error mitigation strategies we applied to reduce the memory error, including dynamical decoupling \cite{viola1999}, logical basis rotation, and Clifford deformation \cite{Dua2024}. With the error mitigation, we were able to keep $p_{\textrm{detect}}(r,t)$ below the target $p=0.2$ value, allowing us to access the critical point of the model in error detected circuits without post-selection. Fig.~\ref{fig:fig4}\textbf{c} shows the main-run reset probabilities at each space-time point, with typical error bars on individual space-time points around $~0.01$ from finite sampling. During this run, resets were injected according to Eq.~(\ref{eq:pinject}) using the $p_{\textrm{detect}}(r,t)$ obtained in the calibration run shown in Fig.~\ref{fig:fig4}\textbf{b}. 
The resulting reset probabilities are more homogeneous and closer to $p=0.2$, however certain points significantly deviate from this target value, which we attribute to imperfections in the calibration due to either statistical errors from finite sampling or drift in the underlying error model between the calibration and main runs (which were always run back-to-back in time).
In an attempt to fix the residual discrepancies between the sampled reset rates and the target one, we implemented an importance-sampling based reweighting scheme, described in Supplement~III.C, that ascribes a weight to each shot based on the ratio of its empirical probability compared to the ideal probability and computes observable values using a weighted average. Upon reweighting, the reset probabilities become homogenous and close to the target value of $p=0.2$ (see Fig.~\ref{fig:fig4}\textbf{d}). The logical observables in Fig.~\ref{fig:fig3} use the reweighting method.

\emph{Discussion---} We demonstrate a scalable use of quantum error detecting codes to adaptively implement a logical simulation of a dissipative quantum circuit that avoids post-selection. Essentially, hardware errors are used as a source of randomness to implement random resets that exist in the dissipative circuit model of interest. This demonstration relied heavily on the high-fidelity, low cross-talk, and fast mid-circuit measurements and resets in the Quantinuum trapped-ion hardware, as well as the classical feed-forward functionality.

Our circuits run on the H2 device appear to break-even for short times, but fail to break-even at the latest time point. We believe that this is the result of two factors: (1) memory errors increasing with circuit depth due to the limited number of parallel gate operations that can be performed in the device and (2) detectable logical errors becoming undetectable logical errors in other code blocks when passing through a non-fault-tolerant interblock gate. Neither of these are fundamental issues, and they can be addressed to improve the scalability of the protocol. Concerning (1), future generations of Quantinuum hardware will have more gate zones allowing parallel two-qubit gate operations, as this is a crucial consideration for scaling up quantum error correcting codes. With more gate zones, one can also be more sparing with qubit-reuse to avoid the linear-in-time scaling of memory error. Concerning (2), using a purely fault-tolerant implementation of each gate, e.g., using magic state injection, will prevent the problematic spreading of undetectable errors, though likely at a large space and time overhead. We leave these improvements, as well as improvements to other details such as our reset injection calibration procedure, for future studies.

Finally, we note that the scalable post-selection-free protocol in this work can be applied to circuits other than the contact process or circuits with random resets. In particular, it can be utilized in circuits containing quantum channels of the form $\mathcal{E}(\rho)=(1-p)\rho + p \rho'$ where $\rho'$ is a density matrix independent of $\rho$. For example, $\rho'=|00\rangle\langle 00|$ corresponds to the two-qubit random reset in the contact process model and $\rho'\propto I$ corresponds to a depolarizing channel. It will be interesting future work to find examples of other post-selection-free error detection protocols that can extend to even larger classes of circuits.

\section*{Acknowledgments}
We acknowledge the entire Quantinuum team for their many contributions towards successful operation of the H2 quantum computer, and Honeywell for fabricating the trap used in this experiment. We thank Chris Self, David Amaro, Aaron Hankin, Michael Mills, Steven Moses, Natalie Brown, Ciaran Ryan-Anderson, Shival Dasu, Matthew DeCross, Reza Haghshenas, Yi-Hsiang Chen, Karl Mayer, and David Stephen for useful discussions.

\bibliography{refs}

\end{document}


\title{Supplemental Material: Error detection without post-selection in adaptive quantum circuits}

\author{Eli Chertkov}
\email{eli.chertkov@quantinuum.com}
\author{Andrew C. Potter}
\author{David Hayes}
\author{Michael Foss-Feig}
\affiliation{Quantinuum, 303 South Technology Court, Broomfield, Colorado 80021, USA}

\maketitle

\beginsupplement

\section{The [[4,2,2]] quantum error detecting code}

The [[4,2,2]] code is the smallest quantum error detection code and has a large encoding rate of $k/n=1/2.$ It has many logical operations, including many fault-tolerant ones, that can be implemented with modest gate overheads. Importantly, there are even non-fault tolerant operations which are particularly resilient to the main noise sources present in the Quantinuum trapped-ion quantum computers \cite{Haghshenas2025}, making them particularly suitable for implementation on such quantum computers.

\subsection{Stabilizers, logical operators, and logical states}

The code has stabilizers 
\begin{align}
S_X = X_1 X_2 X_3 X_4, \quad S_Z = Z_1 Z_2 Z_3 Z_4.
\end{align}
During a logical quantum circuit, the stabilizers are symmetries of the circuit, so that the quantum state is restricted to stay as a $+1$ eigenstate of both operators. By repeatedly measuring if $S_{X/Z}=+1$, we can detect if an error occurs. When acting on a valid logical state, the stabilizers act like identity.

The logical Pauli operators
\begin{align}
&\overline{X}_1 = X_1 X_2 \textrm{ or } X_3 X_4, \quad \overline{Z}_1 = Z_1 Z_3 \textrm{ or } Z_2 Z_4 \nonumber \\ 
&\overline{X}_2 = X_1 X_3 \textrm{ or } X_2 X_4, \quad \overline{Z}_2 = Z_1 Z_2 \textrm{ or } Z_3 Z_4.
\end{align}
commute with the stabilizers and define operations on the logical qubits. Note how there can be multiple equivalent representations of each operator (not all shown), related by multiplication by the stabilizers.

The logical basis states $\ket{\overline{00}},\ket{\overline{01}},\ket{\overline{10}},\ket{\overline{11}}$ are $\pm 1$ eigenstates of the logical operators $\overline{Z}_1,\overline{Z}_2$ and $+1$ eigenstates of the stabilizers $S_X,S_Z$. In terms of the physical qubits, they are
\begin{align}
\ket{\overline{00}} &= \frac{1}{\sqrt{2}} \left(\ket{0000}+\ket{1111}\right) \nonumber \\
\ket{\overline{01}} &= \frac{1}{\sqrt{2}} \left(\ket{0101}+\ket{1010}\right) \nonumber \\
\ket{\overline{10}} &= \frac{1}{\sqrt{2}} \left(\ket{0011}+\ket{1100}\right) \nonumber \\
\ket{\overline{11}} &= \frac{1}{\sqrt{2}} \left(\ket{0110}+\ket{1001}\right). \label{eq:logical_basis}
\end{align}
Note how $\ket{\overline{00}}$ is a GHZ state, while the others are modified GHZ states with two bits flipped.

\subsection{Fault-tolerant logical operations} \label{sec:ft_gates}

\subsubsection{Stabilizer measurement (error detection)}

The stabilizer measurement circuit that we use is
\begin{center}
\scalebox{0.7}{ 
\begin{quantikz}
\lstick[4]{block} & & \targ{} & \ctrl{4} & & & & & & & & \\
& & & & \ctrl{3} & \targ{} & & & & & & \\
& & & & & & \targ{} & \ctrl{2} & & & & \\
& & & & & & & & \ctrl{1} & \targ{} & & \\
\lstick{\ket{0}} & & & \targ{} & \targ{} & & & \targ{} & \targ{} & & & \meter{} \\
\lstick{\ket{0}} & \gate{H} & \ctrl{-5} & & & \ctrl{-4} & \ctrl{-3} & & & \ctrl{-2} & \gate{H} & \meter{} \\
\end{quantikz}
}
\end{center}
which is adapted from Ref.~\onlinecite{Self2024}.

It measures both $S_Z$ and $S_X$, with their results stored into the top and bottom ancilla respectively. This circuit by its construction is fault-tolerant, achieving this fault-tolerance in a similar way as flag-fault-tolerant circuits \cite{Chamberland2018,Chao2020}.

\subsubsection{State preparation (reset)}

We use the $\ket{\overline{00}}$ state preparation (or reset) circuit in Ref.~\onlinecite{Gottesman2016}:

\begin{center}
\scalebox{0.7}{ 
\begin{quantikz}[wire types={n,n,n,n,n}]
\lstick[4]{block} & \lstick{\ket{0}} &\setwiretype{q} & & \targ{} & & \ctrl{4} & \\
&\lstick{\ket{0}} & \gate{H}\setwiretype{q} & \ctrl{1} & \ctrl{-1} & & &\\
&\lstick{\ket{0}} &\setwiretype{q} & \targ{} & \ctrl{1} & & &\\
&\lstick{\ket{0}} &\setwiretype{q} & & \targ{} & \ctrl{1} & &\\
& & & & \lstick{\ket{0}} & \targ{}\setwiretype{q} & \targ{} & \meter{} \\
\end{quantikz}
}
\end{center}

The ancilla qubit is used to guarantee fault-tolerance. If the ancilla measures a $0$ then the state is accepted. We use a repeat-until-success protocol in each logical reset in order to avoid any post-selection overhead associated with discarding bad resets. We repeat each reset up to two times.

\subsubsection{Measurement}

At the end of our circuit, the logical measurements are 
\begin{center}
\scalebox{0.7}{ 
\begin{quantikz}[wire types={q,q,q,q,n,n}]
 \lstick[4]{block} & \ctrl{4}& & & & & & \\ 
 & & & & & \ctrl{4}& &\\
 & & \ctrl{2} & & & & &\\
 & & & & & & \ctrl{2} & \\
 \setwiretype{q}\lstick{\ket{0}} & \targ{} & \targ{} & \meter{} \\
  & & & &\lstick{\ket{0}}  & \targ{}\setwiretype{q} & \targ{} & \meter{} \\
\end{quantikz}
}
\end{center}
to measure $\overline{Z}_1$ and
\begin{center}
\scalebox{0.7}{ 
\begin{quantikz}[wire types={q,q,q,q,n,n}]
\lstick[4]{block} & \ctrl{4}& & & & & & \\ 
 & & \ctrl{3} & & & & &\\
 & & & & & \ctrl{3} & &\\
 & & & & & & \ctrl{2} & \\
 \setwiretype{q}\lstick{\ket{0}} & \targ{} & \targ{} & \meter{} \\
  & & & &\lstick{\ket{0}}  & \targ{}\setwiretype{q} & \targ{} & \meter{} \\
\end{quantikz}
}
\end{center}
to measure $\overline{Z}_2$. These are followed by stabilizer measurements to make them fault-tolerant.

\subsection{Non-fault-tolerant logical gates} \label{sec:non_ft_gates}

Our contact process circuit involves non-Clifford gates in the form of two-qubit controlled-rotation gates $\overline{CR}_x(\theta)$, both within a single [[4,2,2]] code block and between two different code blocks. We implement these gates in a non-fault-tolerant way to avoid the costly overhead of a fault-tolerant implementation, which might involve steps like magic state distillation. As can be seen in Fig.~2\textbf{a}, the circuit we are simulating contains logical $\overline{CR}_x(\theta)$ gates both within single code blocks (with control qubits on either the first or second qubit) and between two code blocks.

The $\overline{CR}_x(\theta)$ gate itself can be decomposed into two two-qubit Clifford gates and two single-qubit non-Clifford rotations:
\begin{align}
\overline{CR}_{x}(\theta)_{1,2}=\bar{R}_{x,2}(\theta/2)(\overline{CZ}_{1,2})\bar{R}_{x,2}(-\theta/2)(\overline{CZ}_{1,2}) \label{eq:crx_decomp}.
\end{align}
Using this decomposition and implementing the FT $\overline{CZ}$ gate and non-FT $\bar{R}_x(\theta)$ gate available in the [[4,2,2]] code \cite{Gottesman2016,Vuillot2017,Chao2018,Harper2019}, we implement the $\overline{CR}_x(\theta)$ gate within a single code block as the following circuits:
\begin{center}
\scalebox{0.7}{ 
\begin{quantikz}
\lstick[4]{block A} & & \phase[open,label style={label
position=above,anchor=north,yshift=+0.2cm}]{+\theta/2} & & \\
& & \ctrl[open]{-1} & & \\
& \gate{S} & \phase[open,label style={label
position=above,anchor=north,yshift=+0.2cm}]{+\theta/2} & \gate{S^\dagger} & \\
& \gate{S} & \ctrl[open]{-1} & \gate{S^\dagger} & \\
\end{quantikz}
}
\end{center}
and
\begin{center}
\scalebox{0.7}{ 
\begin{quantikz}
\lstick[4]{block A} & \phase[open,label style={label
position=above,anchor=north,yshift=+0.2cm}]{+\theta/2} & & & & \\
& & \gate{S} & \phase[open,label style={label
position=above,anchor=north,yshift=+0.2cm}]{+\theta/2} & \gate{S^\dagger} & \\
& \ctrl[open]{-2} & & & & \\
& & \gate{S} & \ctrl[open]{-2} & \gate{S^\dagger} & \\
\end{quantikz}
}
\end{center}
where the top diagram is for $\overline{CR}_x(\theta)_{A_2,A_1}$ (i.e., control qubit $A_2$) and the bottom diagram is for $\overline{CR}_x(\theta)_{A_1,A_2}$, where $A_1,A_2$ are the two logical qubits in code block $A$. The gates with open circles and specified angle $\theta$ are physical $e^{-i\theta X\otimes X/2}$ gates, which can be implemented with a single arbitrary-angle $U_{ZZ}(\theta)=e^{-i\theta Z\otimes Z/2}$ gate that can be executed natively on Quantinuum hardware \cite{Moses2023,Chertkov2023,DeCross2024,Haghshenas2025}. 

Using Eq.~(\ref{eq:crx_decomp}), the FT parallel $\overline{CX}$ gates, FT $(\overline{H} \otimes \overline{H})(\overline{SWAP})$ gates, and non-FT $\bar{R}_x(\theta)$ rotations in the [[4,2,2]] code \cite{Gottesman2016,Vuillot2017,Chao2018,Harper2019}, we implement the $\overline{CR}_x(\theta)$ gate between two code blocks $A$ and $B$ as the following circuit:

\begin{center}
\scalebox{0.7}{ 
\begin{quantikz}
\lstick[4]{block A} & \phase[open,label style={label
position=above,anchor=north,yshift=+0.2cm}]{+\theta/2} & \ctrl{4} & & & \phase[open,label style={label
position=above,anchor=north,yshift=+0.2cm}]{-\theta/2} & \ctrl{4} & & & & \\
& \ctrl[open]{-1} & & \ctrl{5} & & \ctrl[open]{-1} & & \ctrl{5} & & & \\
& & & & \ctrl{3} & & & & \ctrl{3} & & \\
& & & & & & & & & & \\
\lstick[4]{block B} & & \ctrl{0} & & & \phase[open,label style={label
position=above,anchor=north,yshift=+0.2cm}]{-\theta/2} & \ctrl{0} & & & \phase[open,label style={label
position=above,anchor=north,yshift=+0.2cm}]{+\theta/2} & \\
& & & & \ctrl{0} & \ctrl[open]{-1} & & & \ctrl{0} & \ctrl[open]{-1} & \\
& & & \ctrl{0} & & & & \ctrl{0} & & & \\
& & & & & & & & & & \\
\end{quantikz}
}
\end{center}
The gates with closed circles are physical $CZ$ gates. In particular, the above circuit implements the logical gate $(\overline{CR}_x(\theta)_{A_1,B_1})(\overline{CR}_x(\theta)_{B_1,A_1})$.

Since this gate is not FT, errors that occur \emph{before} the gate can spread in a problematic way. In particular, suppose that due to the memory error on the quantum computer, e.g., accumulated during ion transport in a Quantinuum device, a single Pauli-$Z$ is applied to code block B. For simplicity, suppose the angle $\theta=\pi$ so that the circuit is Clifford (but still not FT). In this circuit, the error can spread as shown below:
\begin{center}
\scalebox{0.6}{ 
\begin{quantikz}
\lstick[4]{block A} & \phase[open,label style={label
position=above,anchor=north,yshift=+0.2cm}]{+\pi/2} & \ctrl{4} & & & & \phase[open,label style={label
position=above,anchor=north,yshift=+0.2cm}]{-\pi/2} & & \ctrl{4} & & & & & \push{\color{red}Z} & \\
& \ctrl[open]{-1} & & \ctrl{5} & & & \ctrl[open]{-1} & & & \ctrl{5} & & & & & \\
& & & & \ctrl{3} & & & & & & \ctrl{3} & & & \push{\color{red}Z} & \\
& & & & & & & & & & & & & & \\
\lstick[4]{block B} & \push{\color{red}Z} & \ctrl{0} & & & \push{\color{red}Z}  & \phase[open,label style={label
position=above,anchor=north,yshift=+0.2cm}]{-\pi/2} & \push{\color{red}Y} & \ctrl{0} & & & \push{\color{red}Y} & \phase[open,label style={label
position=above,anchor=north,yshift=+0.2cm}]{+\pi/2} & \push{\color{red}Z} &\\
& & & & \ctrl{0} & & \ctrl[open]{-1} & \push{\color{red}X} & & & \ctrl{0} & \push{\color{red}X} & \ctrl[open]{-1} & & \\
& & & \ctrl{0} & & & & & & \ctrl{0} & & & & & \\
& & & & & & & & & & & & & & \\
\end{quantikz}
}
\end{center}
In this case, the weight-1 Pauli error leads to a detectable error ($Z_1$) on code block B, but an undetectable logical error ($\overline{Z}_1 = Z_1 Z_3$) on code block A.  For these circuits, we think that the logical error is effectively dominated by these memory-error-created undetectable logical errors that spread between code blocks. 

The undetectable logical errors shown above lead to the (memory-error-induced) logical error rate scaling as $p_\mathrm{logical}  \sim C_{\mathrm{logical}} p_{\mathrm{memory}}$ where $C_{\mathrm{logical}}$ is a constant and $p_{\mathrm{memory}}$ is the physical memory error rate. This is clearly not FT, with the physical error rate scaling the same way, $p_{\mathrm{physical}} \sim C_{\mathrm{physical}} p_{\mathrm{memory}}$ where $C_{\mathrm{physical}}$ is a different constant. Due to gate zone limitations and the qubit-reuse in our circuits, both constants grow linearly in time: $C_{\mathrm{physical}} = A_{\mathrm{physical}} t, C_{\mathrm{logical}}=A_{\mathrm{logical}} t$. Since the amount of transport and circuit run-time is much larger for the logical circuit compared to the physical circuit, we expect that $A_\mathrm{logical} > A_\mathrm{physical}$ and that the logical circuit infidelity due to memory error is always worse than the physical circuit infidelity due to memory error. For short circuits, this effect is likely not noticeable because the logical errors in the circuit are dominated by gate errors. However, for the deepest circuits we consider, such as $t=7$ in Fig.~3, the dominant error source is likely the memory error, where the logical circuits perform worse.

In principle, this problematic scaling can be solved by making the circuit FT or by performing error detection frequently enough so that the amount of memory error that leads to undetectable logical errors is low enough. 

\section{The two-qubit-reset floquet quantum contact process}

In this work, we consider a variant of the floquet quantum contact process model studied in Ref.~\onlinecite{Chertkov2023} with two-qubit random resets instead of one-qubit random resets. Given this difference, we perform numerical simulations to assess whether this new model possesses the same critical behavior as the original model. Similar to Ref.~\onlinecite{Chertkov2023}, we perform numerical simulation on a classical limit of the model, which can be efficiently simulated for large system sizes and times, and on the quantum model.

While many observables can be studied to understand the critical behavior of the model, for simplicity we focus on the single observable $\langle N_R(t)\rangle = \sum_{r\geq 0} \langle n(r,t)\rangle$, the total number of active sites (qubits in the $\ket{1}$ state) on the right-half of the system. We always initialize our time evolution as described in the main text, with a single active site in the center of the chain. When we time-evolve to a time $t$, we always use a system size large enough to contain the entire causal cone of the circuit, as shown in Fig.~2\textbf{a}. This effectively makes our simulations in the thermodynamic limit and removes any finite-size effects, though there are still finite-time effects.

For both the classical and quantum limits, we locate the phase transition by analyzing the effective exponent for the $\langle N_R(t)\rangle$ observable. The effective exponent of observable $O$ is defined as 
\begin{align}
\delta_O(t) = \frac{\log [O(t+dt)/O(t)]}{\log [(t+dt)/t]} = \frac{\log [O(t+dt)] - \log[O(t)]}{\log [t+dt] - \log[t]}. \label{eq:eff_exp}
\end{align}
When an observable scales as a power-law with exponent $\delta'$ ($O(t)=O_0 t^{\delta'}$), $\delta_O(t)=\delta'$ is a constant. At the critical point of the model, we expect power-law scaling of $O(t)$ and $t$-independence of $\delta_O(t)$ at late times. Therefore, when plotting $\delta_O(t)$ versus $t$ curves for many fixed values of $p$, we can identify the critical point $p_c$ by the location where the $\delta_O(t)$ curves cross.

\subsection{Classical (fully dephased) model} \label{sec:classical_point_sims}

In the one-qubit-reset model studied in Ref.~\onlinecite{Chertkov2023}, the model was analyzed in the classical limit defined by taking $\theta \rightarrow \pi/2$. In this limit, the circuit becomes a Clifford circuit. Moreover, the circuit dynamics on bit-strings generates no entanglement, making the dynamics an efficiently simulatable classical stochastic process on bit-strings. A similar limit exists in the two-qubit-reset model studied here. However, from our numerical simulations we find that the Clifford point of this model always reaches the absorbing state for all values of $p$. Evidently, there is a hidden symmetry at the Clifford point that destroys the absorbing state transition, making it non-representative of the quantum model. 

Therefore, we instead study a different classical limit of the quantum two-qubit-reset model. We study a \emph{fully-dephased} limit of the quantum circuit, where after each quantum operation a full dephasing channel is applied to the qubits. This limit can be applied to any quantum circuit to construct a classical stochastic process from it. The classical stochastic process describes the dynamics of bit-strings subject to the following rule: after each quantum gate is applied, the qubits in those gates are measured in the $Z$-basis. 

\begin{figure}
    \centering
    \includegraphics[width=0.5\textwidth]{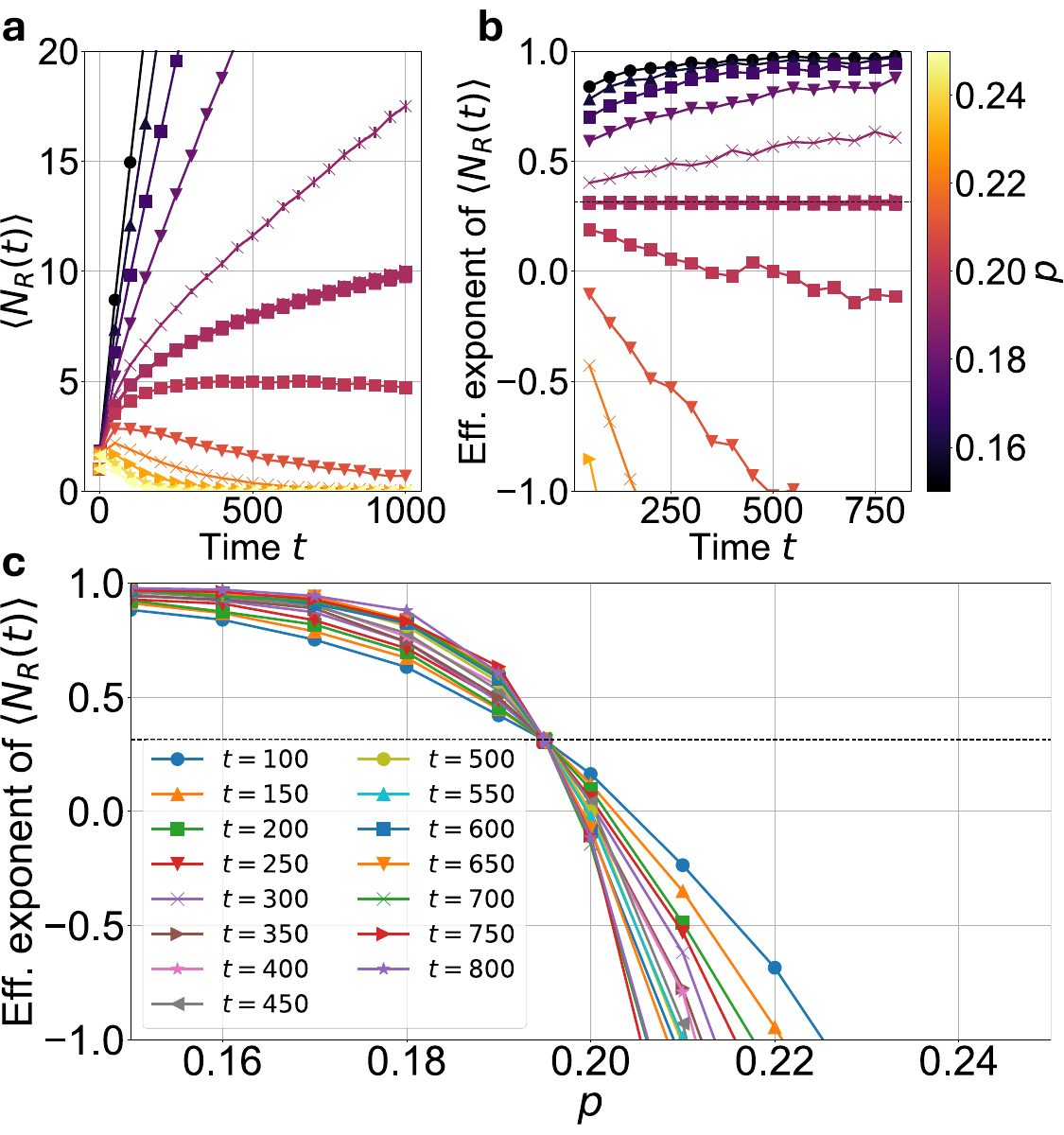}
    \caption{\textbf{a} The number of active sites on the right-half of the system versus time $t$ for the \emph{classical (fully dephased)} model, obtained from large-scale classical numerics. Different colors correspond to different values of reset rate $p$. \textbf{b} The effective exponent Eq.~(\ref{eq:eff_exp}) corresponding to the observable. \textbf{c} The effective exponent versus reset rate $p$ for different fixed values of $t$, showing a crossing at $p_c \approx 0.195$. Horizontal dashed lines correspond to the directed percolation exponent $\Theta\approx 0.314$.}
    \label{fig:fig_classical_point}
\end{figure}

We perform large-scale numerical simulations classical limit of the two-qubit model to times up to $t=1000$ for many values of $p$. In Fig.~\ref{fig:fig_classical_point}\textbf{a}, we show the right-half number of active sites versus time, showing qualitatively the expected decay of active sites at small $p$, the ballistic growth at large $p$, and the power-law growth near a critical $p_c$. Fig.~\ref{fig:fig_classical_point}\textbf{b} shows the effective exponent of the observable, which is constant at the critical $p_c$.  The crossings of the effective exponent curves is displayed in Fig.~\ref{fig:fig_classical_point}\textbf{c}, which locates the transition at $p_c\approx0.195$ and confirms that the scaling of the observables matches the expected DP scaling: $\langle N_R(t)\rangle \sim t^\Theta$ for $\Theta\approx 0.314$. For these simulations, we use a large value of $dt=200$ in Eq.~(\ref{eq:eff_exp}) to reduce the statistical errors on the effective exponents. The observables are averaged over $11,200$ ($11,200,000$) samples generated from the stochastic process for $p$ far from $p_c$ (near $p_c$).

\subsection{Quantum model} \label{sec:quantum_point_sims}

We perform a similar numerical analysis on the quantum model, though at small scales given the difficulty of numerically simulating the quantum behavior. We simulate the model using an exact simulation, where the full open-system evolution is captured exactly using a density matrix. To reach as late of times as possible, we use the qubit-reuse technique \cite{Chertkov2023,DeCross2022} to smartly measure and reset qubits in the numerical simulation, reducing the required number of qubits and therefore the required size of the density matrix needed to perform the calculations. 

\begin{figure}
    \centering
    \includegraphics[width=0.5\textwidth]{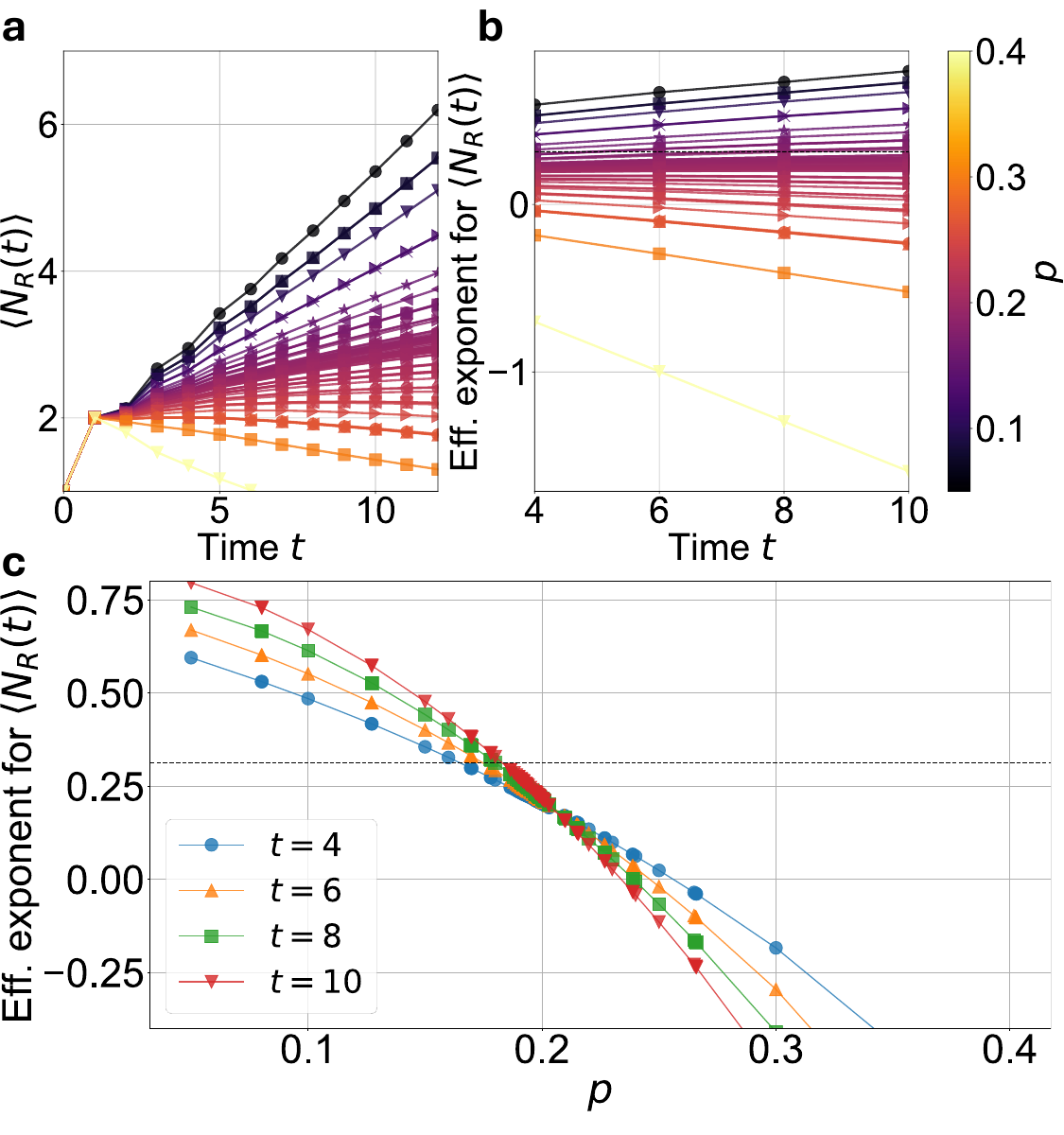}
    \caption{\textbf{a} The number of active sites on the right-half of the system versus time $t$ for the \emph{quantum} model, obtained from small-scale classical density-matrix numerics. Different colors correspond to different values of reset rate $p$. \textbf{b} The effective exponent Eq.~(\ref{eq:eff_exp}) corresponding to the observable. \textbf{c} The effective exponent versus reset rate $p$ for different fixed values of $t$, showing a crossing at $p_c \approx 0.2$. Horizontal dashed lines correspond to the directed percolation exponent $\Theta\approx 0.314$.}
    \label{fig:fig_quantum_point}
\end{figure}

The exact numerical simulations up to $t=12$ are displayed in Fig.~\ref{fig:fig_quantum_point}. The right-side number of active sites and its effective exponents are shown in Fig.~\ref{fig:fig_quantum_point}\textbf{a}~and~\textbf{b}, respectively. The effective exponents calculated here use $dt=2$ in Eq.~(\ref{eq:eff_exp}), given the limited amount of time evolution available. The crossing of the effective exponent curves are shown in Fig.~\ref{fig:fig_quantum_point}\textbf{c}. From the crossings, we roughly estimate that the critical point for the quantum model is at $p_c \approx 0.2$. We note that the value of the effective exponent is not that close to the $\Theta=0.314$ value at $p_c=0.2$. While larger-scale numerics will be needed to confirm this, we speculate that this is a finite-time effect, since similar behavior was observed in Ref.~\onlinecite{Chertkov2023}.

Ultimately, our numerical simulations appear consistent with the quantum two-qubit model undergoing a DP transition at $p_c\approx 0.2$. Note that this is a lower reset value than that of the critical point in the one-qubit-reset model, $p_c^{\textrm{(1Q reset)}}\approx 0.3$, due to the larger effect of each individual reset.

\section{Performance on H2 quantum computer}

In this section, we overview the implementation details of the contact process circuits executed on the H2 quantum computer, particularly as they relate to mitigating errors from hardware noise.

\subsection{Qubit-reuse}

\begin{figure}
    \centering
    \includegraphics[width=0.5\textwidth]{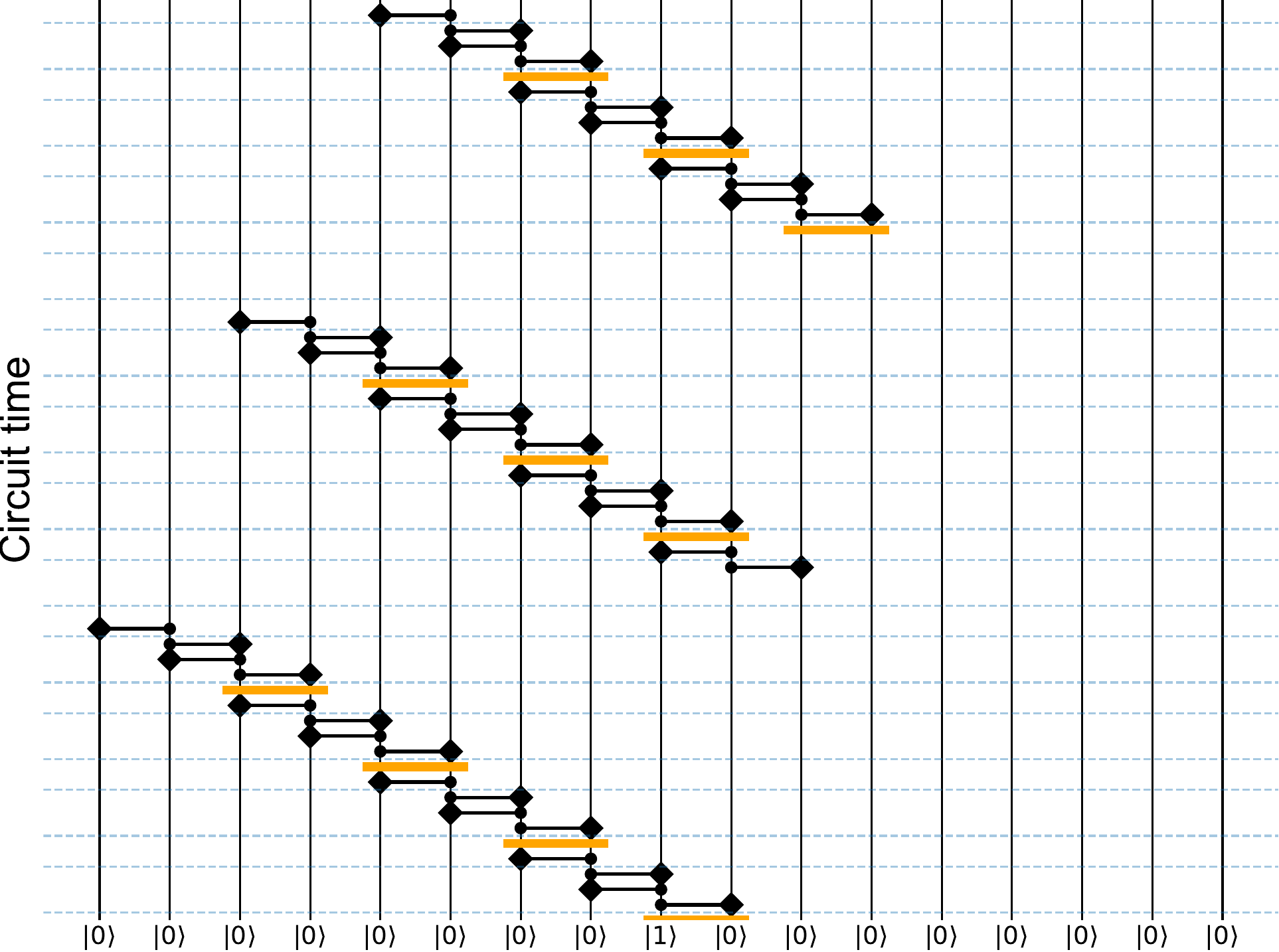}
    \caption{A schematic of the logical circuit, showing how the logical operations are executed in time when performing qubit reuse. Not shown are mid-circuit logical measurements and how logical code blocks are reset and reused as other code blocks. The horizontal dashed lines show locations at which dynamical decoupling pulses are applied.}
    \label{fig:fig_qubit_reuse}
\end{figure}

The physical and logical circuit implementations of the contact process involve the same qubit-reuse as implemented and described in Ref.~\onlinecite{Chertkov2023}. Fig.~\ref{fig:fig_qubit_reuse} shows how logical operations in the circuit occur in time, highlighting the serial nature of qubit-reuse circuits. Given the non-trivial gate decompositions of each logical operation in the circuit (discussed above in the supplement), there can be significant ion transport and ion idling in between logical operations that naively would appear to be successive in the original circuit shown in Fig.~2. The transport and idling leads to an accumulation in memory errors on the qubits. In these qubit-reused circuits, the amount of time elapsed between qubit reuse ``slices'' scales as $\sim t$, which means that the larger $t$ circuits are particularly affected by memory errors.

\subsection{Reducing the effects of memory errors}

We use three different strategies to mitigate the effects of coherent and incoherent memory errors in our logical circuit experiments.

\subsubsection{Circuit-level dynamical decoupling}

We use a crude form of dynamical decoupling (DD) \cite{viola1999} in an attempt to coherently eliminate coherent dephasing on the qubits during idling and transport. We use a variant of the technique described in Ref.~\onlinecite{Yamamoto2024}, for performing DD on a logical circuit. In this technique, we make use of the fact that since $XXXX$ is a stabilizer of the [[4,2,2]] code we can freely apply it to a code block without affecting the logical dynamics. We insert $\prod_j X_j$, which is a product of the $X$ stabilizers on all code blocks, throughout our circuit after most logical operations so that $X$ pulses occur roughly evenly spaced in time. In particular, we insert the $X$ pulses at the locations of the dashed horizontal lines shown in Fig.~\ref{fig:fig_qubit_reuse}. Empirically, we find that the spacing is roughly uniform and from numerical simulations observe a significant improvement in circuit performance with the DD pulses. 

We also reduce the impact of coherent single-qubit gate errors by alternating the phase of the $X$ gates performed during DD, as was done in Ref.~\onlinecite{Haghshenas2025}. This is done by alternating the phase of the native $U_{1q}(\theta,\phi)$ gate used to implement each $X$ gate (odd $X$ pulses are implemented with $U_{1q}(\pi,0)=-iX$ and even $X$ pulse with $U_{1q}(\pi,\pi) = +iX$). The alternating phase leads to coherent cancellation of single-qubit gate errors.

\subsubsection{Decoherence-free subspaces}

Decoherence-free subspaces (DFS) are subspaces of Hilbert space that are unaffected by uniform dephasing \cite{Lidar2003,Lidar2014}. For example, all bit-strings with a fixed number of $1$s and $0$s, such as $\ket{0011},\ket{1010},\ldots$, span a DFS and are unaffected (up to a global phase) by the uniform dephasing operator $D=\prod_j e^{-i\theta Z_j}$. For a single code block of the [[4,2,2]] code, three out of four of the logical basis states $\ket{\overline{s}}$ for $s=01,10,11$ (see Eq.~(\ref{eq:logical_basis})) form a DFS and are completely unaffected by uniform dephasing $D\ket{s}=\ket{s}$. However, in contrast, the $\ket{\overline{00}}$ logical state is a physical GHZ state, which is maximally sensitive to uniform dephasing: $D\ket{\overline{00}}\propto\frac{1}{\sqrt{2}}\left(\ket{0000}+e^{8i\theta}\ket{1111}\right)$.

Since we expect that coherent dephasing in Quantinuum devices is approximately uniform in space and time and since the contact process circuits are biased to be in the $\ket{\overline{00}}$ state due to the random resets, we want the $\ket{\overline{00}}$ state to be in a DFS. We are able to do this by conjugating the logical resets and gates in our circuits by $\overline{X}_1 \overline{X}_2=X_2 X_3$. This essentially amounts to a permutation among the logical basis states to the new logical basis:
\begin{align}
\ket{\overline{00}} &\rightarrow\ket{\widetilde{11}}= \frac{1}{\sqrt{2}} \left(\ket{0000}+\ket{1111}\right) \nonumber \\
\ket{\overline{01}} &\rightarrow\ket{\widetilde{10}}= \frac{1}{\sqrt{2}} \left(\ket{0101}+\ket{1010}\right) \nonumber \\
\ket{\overline{10}} &\rightarrow\ket{\widetilde{01}}= \frac{1}{\sqrt{2}} \left(\ket{0011}+\ket{1100}\right) \nonumber \\
\ket{\overline{11}} &\rightarrow\ket{\widetilde{00}}= \frac{1}{\sqrt{2}} \left(\ket{0110}+\ket{1001}\right). \label{eq:new_logical_basis}
\end{align}

In the circuits run in our experiments, we use this new logical basis so that the logical $\ket{\widetilde{00}}$ state is a DFS state and therefore robust to uniform dephasing.

\subsubsection{Clifford deformation}

Clifford deformation is a technique for suppressing logical memory error in a quantum error correction (QEC) code with a biased memory error channel \cite{Dua2024}. It amounts to conjugating the QEC code by single-qubit Clifford gates, potentially making the code non-CSS. This has the effect of changing the stabilizers measured during syndrome extraction. If the memory error noise model is biased, e.g., towards $Z$ noise as is the case in Quantinuum hardware, then having more stabilizers contain more $X$ and $Y$ Paulis can help catch a larger fraction of the memory errors. Here we apply this principle to the [[4,2,2]] code.

Suppose that during the transport in between two successive logical operations on a single code block, memory error accumulates so that a weight-1 $Z$ error happens on any of the $4$ qubits with equal probability $p_\mathrm{memory}$. Any two weight-1 $Z$ errors in the [[4,2,2]] code causes a logical error (e.g., $Z_1$ and $Z_4$ cause a $Z_1 Z_4 = \overline{Z}_1 \overline{Z}_2$ error). Counting up all possible ways this can happen, we see that the memory-error-induced logical error rate is $\binom{4}{2}p_\mathrm{memory}^2=6p_\mathrm{memory}^2$. 

Now instead, suppose that before and after the transport is initiated the code block is conjugated by the Clifford operator $I\otimes I \otimes H \otimes (HS)$, so that weight-1 $Z$ Paulis occuring during transport transform to $Z_1 \rightarrow Z_1, Z_2 \rightarrow Z_2, Z_3 \rightarrow X_3, Z_4 \rightarrow Y_4$. Now if we consider all error events with two weight-1 processes, we see that many of the processes (e.g., $Z_1$ and $Z_3$ cause a $Z_1 X_3$ error that triggers $S_X=-1,S_Z=-1$) are now detectable errors. With the Clifford deformed circuit, the only way a logical error can occur is if $Z_1$ and $Z_2$ errors occur, which happens at a rate of $p_\mathrm{memory}^2$, $6\times$ lower than the original circuit. 

In our experiments, we use Clifford deformation, the DFS basis transformation, and DD. In principle, Clifford deformation can potentially reduce the effectiveness of the DFS basis transformation. However, in numerical simulations we saw a slight improvement when using both together, so we used both in the experiments.

\subsection{Reset reweighting scheme} \label{sec:reweighting}

As described in the main text, in the adaptive circuit protocol the final reset rate measured at each space-time point is not necessarily uniform in space and time as defined in the model, due to statistical and systematic errors in the reset injection protocol. In an attempt to correct for this non-uniformity, we implement a reweighting protocol inspired by importance sampling.

Consider a space-time point $(r,t)$ in the circuit. In the main run of the circuit, we have an empirically measured reset rate $p(r,t)$ that includes contributions of detected and injected resets and differs from the ideal value $p$. For each shot $s$, we have a record of the local density observable $n_s(r,t)$ and whether a reset was observed $m_s(r,t)$ at the space-time point. Without any reweighting, the local density is estimated as $\langle n(r,t)\rangle \approx \frac{1}{M}\sum_{s=1}^M n_s(r,t)$ for $M$ shots. With reweighting, we measure the local density as a weighted average
\begin{align}
\langle n(r,t)\rangle_{\mathrm{reweighted}} = \frac{1}{M}\sum_{s=1}^M w_s n_s(r,t) \label{eq:n_reweighted}
\end{align}
where the weight for a shot $s$ is defined as
\begin{align}
w_s &\equiv \frac{P_\mathrm{ideal}(s)}{P_\mathrm{empirical}(s)} \nonumber \\
&=\prod_{(r',t')}\left(\frac{p}{p(r',t')}\right)^{m_s(r',t')} \left(\frac{1-p}{1-p(r',t')}\right)^{1-m_s(r',t')}. \label{eq:weights}
\end{align}
The weights in Eq.~(\ref{eq:weights}) are a ratio of the ideal probability of obtaining the reset pattern in shot $s$ for a uniform reset rate $p$ over the empirical probability of obtaining the pattern observed in the $M$-shot dataset. We assume here that resets are uncorrelated in space and time so that the probability distribution factorizes into a product of Bernoulli probabilities for each space-time point. In the limit of large $M$ and correctly estimated $p(r,t)$, the reweighted expectation values of Eq.~(\ref{eq:n_reweighted}) should equal to the ideal expectations with uniform reset rates:
\begin{align*}
\langle n(r,t)\rangle_{\mathrm{reweighted}} &= \frac{1}{M}\sum_{s=1}^M w_s n_s(r,t)\\
&\approx \sum_s \frac{P_\mathrm{ideal}(s)}{P_\mathrm{empirical}(s)} P_\mathrm{empirical}(s) n_s(r,t) \\
&=\sum_s P_\mathrm{ideal}(s) n_s(r,t) = \langle n(r,t) \rangle_\mathrm{ideal}.
\end{align*}
 
\bibliography{refs}
